\def\lte{\lower 0.5ex\hbox{${}\buildrel<\over\sim{}$}}
\def\gte{\lower 0.5ex\hbox{${}\buildrel>\over\sim{}$}}

\documentstyle[12pt,aasms,twoside]{article}
\begin{document}

\title{ON THE LOCATION OF THE ACCELERATION AND }
\title{EMISSION SITES IN GAMMA-RAY BLAZARS}\par \vskip 0.2in

\author{Charles D. Dermer$^1$ and Reinhard Schlickeiser$^2$}

\begin{abstract}
Compton scattering of external radiation by
nonthermal particles in outflowing blazar jets is dominated by accretion-disk
photons rather than scattered radiation to distances $\sim 0.01-0.1$ pc from
the
central engine for standard parameters, thus clarifying the limits of validity
of
the model by the present authors and the model of Sikora, Begelman, \& Rees. On
the basis of contemporaneous Ginga X-ray and EGRET gamma-ray observations, we
estimate the radius of 3C 279's gamma-ray photosphere to be smaller than
estimated
by Blandford.  There is thus no need to require that the acceleration and
emission
sites of gamma-ray blazars to be located farther than $\sim 10^{2-3}$
gravitational radii from the central engine. We argue that lineless BL Lac
objects, rather than quasars, are more likely to be detected in the TeV energy
range. \end{abstract}


\vskip0.2in \noindent $^1$E. O. Hulburt Center for Space Physics, Code 7653,
Naval Research Laboratory,
\linebreak Washington, DC 20375-5352
\vskip0.01in
\noindent{$^2$Max-Planck-Institut f\"ur Radioastronomie, Auf dem H\"ugel 69,
D-53121 Bonn, Germany}

\clearpage

\section{INTRODUCTION}

Gamma-ray observations of blazars provide a new probe of particle
energization by supermassive black holes. The Energetic Gamma Ray Experiment
Telescope (EGRET) on the Compton Observatory has now detected and identified 23
extragalactic sources of $\sim 100$ MeV - 1 GeV emission (for reviews, see
Fichtel et al. 1992; Dermer \& Schlickeiser 1992). The EGRET sources display
blazar properties, which include flat-spectrum radio emission associated with a
compact core, apparent superluminal motion, rapid optical variability and large
optical polarization. Although the origin of the gamma-ray emission is a
subject
of considerable theoretical controversy, a consensus has developed that the
gamma-ray emission is produced in association with the radio jets. Dermer,
Schlickeiser, \& Mastichiadis (1992) proposed that the high energy emission is
produced when energetic electrons in the outflowing radio jets Compton-scatter
external radiation emitted by the accretion disk. Because of the angular
distribution of the radiating electrons in the comoving fluid frame, soft
photons
entering directly from behind are preferentially scattered near the
superluminal
direction. Even when the photons enter from a large angle with respect to the
jet
axis, detection of sources displaying superluminal motion is favored (Dermer \&
Schlickeiser 1993; hereafter DS).

In another blazar model also invoking the Compton scattering of external
photons,
Sikora, Begelman, \& Rees (1993a,b) propose that UV photons scattered by
difffuse
gas or emission line clouds surrounding the central engine provide a more
important soft photon source than accretion disk photons because of the strong
dependence of the scattered flux on the angle that the soft photons make with
respect to the radio axis. Indeed, DS show that the scattered  flux in the
direction that the emission is most intense varies as $\Gamma^2$ for photons
entering from behind, and as $\Gamma^6$ for photons entering from the side,
where
$\Gamma$ is the bulk Lorentz factor of the outflowing jet. However, Sikora et
al.
do not justify their assumption that accretion disk photons are unimportant.
Here
we quantitatively characterize the regime where scattered photons are more
important than accretion disk photons. We find that the scattered photons
dominate rather far from central engine, namely $\gte$ 0.01-0.1 pc for a $10^8
M_\odot$ black hole surrounded by scattering clouds with a mean Thomson
scattering depth $\tau_{sc} \sim 0.01$. Thus one can neglect the accretion-disk
photons only if one maintains that the acceleration and emission site is found
many thousands to tens of thousands gravitational radii from the supermassive
black hole.

In a study complementary with the Sikora et al. model, Blandford (1993) argues
that GeV gamma rays will be strongly attenuated by scattered photons unless
they
are emitted farther than $\sim 0.1$ pc from the central engine. We also perform
this estimate and show that photon attenuation is not a serious problem for 3C
279 unless $\tau_{sc}\sim$ unity within $\sim 0.1$ pc, which is extremely
unlikely. TeV gamma rays could, however, be seriously attenuated by
central-source photons scattered by gas surrounding the jet. If BL Lac objects
(unlike quasars) lack emission line clouds, as is commonly argued in view of
the
weak or absent emission lines in their spectra (e.g., Lawrence 1987), then TeV
photons in these objects can escape unattenuated, and such objects should be
preferred candidates for VHE gamma-ray monitoring. We note, however, that many
BL
Lac objects, including BL Lac itself, do have moderately strong emission lines,
so this argument principally applies to the lineless BL Lac objects.

\section{ELECTRON ENERGY-LOSS RATES IN OUTFLOWING PLASMA JETS}

We can most simply determine the relative importance of radiation fields by
calculating the energy loss rate of energetic electrons in the comoving frame
of
the relativistically outflowing plasma. Here we provide a simplified derivation
of
the gyrophased-averaged electron energy loss rates. For a more detailed
treatment, see DS.

Let $m_ec^2 u_{ph}(\epsilon,\Omega)d\epsilon d\Omega$ represent the total
energy
density of photons with dimensionless energy $\epsilon = h\nu/m_ec^2$ between
$\epsilon$ and $\epsilon + d\epsilon$ that are directed into solid angle
$d\Omega$ in the direction $\Omega$. The quantity
$u_{ph}(\epsilon,\Omega)/\epsilon^3$ is invariant (Rybicki and Lightman 1979).
We
denote photon angles and energies in the stationary (accretion-disk) frame by
asterisks. Quantities in the comoving fluid frame will be unstarred. The
relevant
Lorentz transformation equations are given by $\epsilon^* =
\Gamma\epsilon(1+\beta_\Gamma\mu)$ and $\mu^* =
(\mu+\beta_\Gamma)/(1+\beta_\Gamma\mu)$, where $\cos^{-1}\mu^*$ is the angle a
photon makes with respect to the jet axis. Assuming azimuthal symmetry,

$$u_{ph}(\epsilon,\mu) = {\epsilon^3\over \epsilon^{*3}}\;
u_{ph}^*[\epsilon^*(\epsilon,\mu),\mu^*(\mu)]\;.\eqno(1)$$

If photons are isotropic in the stationary frame,

$$u_{ph}^*(\epsilon^*,\mu^*) = {u_i^*(\epsilon^*)\over 2}\;.\eqno(2)$$

\noindent If photons enter directly from behind,

$$u_{ph}^*(\epsilon^*,\mu^*) = u_b^*(\epsilon^*)\;\delta(\mu^*-1)\;.\eqno(3)$$

\noindent The total energy density of soft photons in the comoving frame is
given
by

$$u_{ph} = \int_{-1}^1 d\mu \int_0^\infty d\epsilon\;
{u_{ph}^*[\Gamma\epsilon(1+\beta_\Gamma\mu),
(\mu+\beta_\Gamma)/(1+\beta_\Gamma\mu)]\over
\Gamma^3(1+\beta_\Gamma\mu)^3}\;.\eqno(4)$$

\noindent Substituting equations (2) and (3) into equation (4) and performing
the
elementary integrals gives

$$u_{ph}^i = u_{ph,i}^* \Gamma^2 (1+\beta_\Gamma^2/3)\;,\eqno(5)$$

\noindent and

$$ u_{ph}^b = {u_{ph,b}^* \over \Gamma^2 (1+\beta_\Gamma)^2}\;,\eqno(6)$$

\noindent where $u_{ph,i}^*$ and $u_{ph,b}^*$ represent the total photon energy
densities measured in the stationary frame for isotropic photons and photons
entering from behind, respectively.

As is well-known, the energy-loss rate of a relativistic electron with Lorentz
factor $\gamma$ in an isotropic radiation field with energy density $u_{ph}$ is
given in the Thomson limit by

$$-\dot\gamma = {4\over 3}c\sigma_T u_{ph} \gamma^2\;.\eqno(7)$$

\noindent We perform a gyrophase averaging of the electron energy-loss rate in
the outflowing fluid frame. This is equivalent to calculating the energy-loss
rate of an electron in an isotropic radiation field with  energy density given
by
equations (5) or (6), assuming all scattering takes place in the Thomson
regime.
Thus the gyrophase-averaged electron energy-loss rate of an electron in a
relativistically outflowing fluid immersed in a radiation field that is
isotropic
is the stationary frame is given by

$$-\dot\gamma_i = {4\over 3}c\sigma_T u_{ph,i}^* \gamma^2\Gamma^2
(1+\beta_\Gamma^2/3)\;.\eqno(8)$$

\noindent If the photons impinge on the outflowing fluid directly from behind,
the
gyrophase-averaged loss rate is given by

$$-\dot\gamma_b = {4\over 3}c\sigma_T u_{ph,b}^* \;{\gamma^2\over\Gamma^2
(1+\beta_\Gamma)^2}\;.\eqno(9)$$

\noindent These results were originally derived by DS.

Letting $L_{ad}$ represent the total luminosity emitted by the accretion disk,
we
can write

$$u_{ph,i}^* \cong {L_{ad}\tau_{sc}\over 4\pi R_{sc}^2 c m_ec^2}\;,\eqno(10)$$

\noindent where $\tau_{sc}$ is the mean scattering depth of the electron
scattering cloud, assumed to be spherically symmetric about the central source,
and $R_{sc}$ is the radial extent of the scattering cloud. If we make the
assumption that the central source is sufficiently compact that directly
produced
photons impinge on the outflowing plasma jets almost directly from behind, then

$$u_{ph,b}^* \cong {L_{ad}\over 4\pi z^2 c m_ec^2}\;,\eqno(11)$$\

\noindent where z is the distance of the plasma blob along the jet axis from
the
central source.

A direct comparison of the resulting energy loss rates shows that, in the limit
$\beta_\Gamma\rightarrow 1$, $|\dot\gamma_i/\dot\gamma_b|> 1$ implies

$$z{\rm (pc)} \gte 0.043 \; {R_{sc}({\rm pc})\over
\Gamma_{10}^2\tau_{-2}^{1/2}}\; ,\eqno(12)$$

\noindent where $\Gamma = 10\Gamma_{10}$ and the scattering depth $\tau_{sc} =
0.01\tau_{-2}$. In this point source approximation for the accretion-disk
radiation, we see that scattered radiation dominates at $z\lte 0.01 - 0.1$ pc
only when $\tau_{sc}\gg 0.01$, $R_{sc}\ll 1$ pc, or $\Gamma \gg 10$.

The previous estimate does not, however, take into account the physical extent
of
the disk. Due to the strong angle-dependence of the scattered flux on the angle
that the soft photons make with respect to the axis of the jet, photons
produced
far out in the disk that enter the jet from the side can make a more important
contribution to the loss rate than the luminous emission emitted from the
innermost regions of the supermassive black hole. DS have derived the
gyrophase-averaged electron energy loss rate when the external photon source is
a
cool outer blackbody. Here we give an illustrative derivation of this result.

Photons which impinge at large angles with respect to the jet axis, namely
those
produced at accretion-disk radii $R\cong z$, dominate the electron
energy-loss rate in the comoving fluid frame. The energy density of these
photons
in the accretion-disk frame can be approximated by

$$u_{ph,ad}^* \approx {L(R\cong z)\over 4\pi (z^2+R^2) c m_ec^2}\;.\eqno(13)$$

\noindent The luminosity emitted within a decade of radii about R is
approximately given by

$$L(R\cong z) \cong L_{ad}\;({R_g\over z})\;,\eqno(14)$$

\noindent where $L_{ad}$ is the total luminosity radiated by the accretion
disk,
and $R_g \equiv GM/c^2 = 1.48\times 10^{13} M_8$ cm is the gravitational radius
of a black hole of mass $10^8 M_8$ solar masses.

Photons produced at $R\cong z$ can be treated as an isotropic photon source,
since
in either case the photons enter the relativistic moving blob at large angles.
Replacing $u_{ph,i}^*$ with $u_{ph,ad}^*$ in equation (8) gives, in the limit
$\beta_{\Gamma}\rightarrow 1$,

$$-\dot\gamma_{ad} \approx 8.6\times
10^{-7}\;{L_{tot}(\rm{ergs~s^{-1}})M_8\over
z^3(\rm{cm})}\;\gamma^2\Gamma^2\;.\eqno(15)$$

\noindent This agrees favorably with the result derived in detail by DS [from
equations (4.2), (4.3), (5.4) and (5.7)], namely

$$-\dot\gamma_{ad} \cong 7.4\times 10^{-8}\;{L_{tot}(\rm{ergs~s^{-1}})M_8\over
\epsilon_f\;z^3(\rm{cm})}\;\gamma^2\Gamma^2\;.\eqno(16)$$

\noindent In equation (16), the radiation effeciency $\epsilon_f = 0.057-0.25$
for a Schwarzschild black hole, depending on the assumed value of the angular
momentum deposited at the inner edge of the accretion disk.

A direct comparison of equations (8) and (15) shows that the electron
energy-loss
rate from scattered photons with energy density described by equation (10)
dominates the loss rate from accretion disk photons only at distances

$$z(\rm{pc}) \gte 0.06\; {M_8^{1/3}\;R_{sc}^{2/3}({\rm pc})\over
\tau_{-2}^{1/3}}\;.\eqno(17)$$

\noindent Note that equation (17) is independent of $\Gamma$.

Sikora et al. (1993a,b) deduce that the emission site in their model is located
between $\approx$ 0.01 and 0.1 pc from the central source. The minimum
distances
at which the accretion-disk radiation can be neglected is the larger of
equation
(12), representing losses from the luminous emission made close to the central
source, and equation (17), representing losses from the lower luminosity
radiation emitted by the outlying portions of the disk.  This assumption can be
justified if there are dense scattering clouds ($\tau_{-2}\gg 1$) much closer
than a parsec from the central source. Ginga observations (e.g., Turner et al.
1989) of quasars, however, do not support the existence of clouds with $N_H \gg
10^{22}$ cm$^{-2}$ in the line-of-sight, and observations of 3C 279 (Makino et
al. 1989) imply values of $N_H<4\times 10^{21}$ cm$^{-2}$ in June 1987 and July
1988. Ginga X-ray observations made during the Viewing Period 3 gamma-ray flare
of 3C 279 are consistent with galactic absorption, i.e., N$_H < 10^{21}$
cm$^{-2}$ (Makino 1993). The acceleration and emission site in the model of DS
is
located between $\sim 10^{-3}$ and $10^{-2}$ pc from the black hole, so the
neglect of the scattered radiation is justified.

\section{PHOTON ATTENUATION FROM SCATTERED RADIATION}

Blandford (1992) has recently argued that GeV radiation produced within $\sim
10^{17}$ cm of the central engine will be pair attenuated in collisions with
scattered central-source emission when $\tau_{-2}\sim 1$ within $\sim 0.1$ pc.
If
the photon spectral index $\alpha > 1$ for the soft photons, higher energy
gamma
rays are even more strongly attenuated. The absence of any clear pair
attenuation
cutoff in the spectra of gamma-ray blazars (Fichtel et al. 1992) would
therefore
imply that the GeV emission is made at  rather large distances ($\gte 0.1$ pc)
from the galactic nucleus. A further consequence would be that the variability
time scale increases with increasing gamma-ray energy. We reconsider this
estimate and show, on the basis of contemporaneous Ginga observations, that the
1
GeV pair attenuation depth $\tau_{\gamma-\gamma}\ll 1$ for likely parameters of
3C 279.

The photon-photon pair attenuation optical depth between heights $z_i$ and $z$
for a photon with dimensionless energy $\epsilon_1 \equiv h\nu_1/m_ec^2$ is
given
by

$$\tau_{\gamma-\gamma}(\epsilon_1) = {1\over 2}\;\int_{z_i}^z dz'\int_{-1}^1
d\mu(1-\mu)\int_{{2\over \epsilon_1(1-\mu)}}^\infty
d\epsilon\;\sigma_{\gamma-\gamma}(\epsilon_1,\epsilon,\mu)
n_{ph}(\epsilon,\mu;z') \eqno(18)$$

\noindent (e.g., Gould \& Schr\'eder 1967; for corrections, see Brown,
Mikaelian
\& Gould 1973), where $z_i$ is the height at which a photon is ejected radially
outward and $n_{ph}(\epsilon,\mu;z')$ is the angle- and energy-dependent photon
density at height $z'$ above the central engine.

If the scattering gas is assumed to be spherically distributed about the
central
nucleus (note that a flattened disk distribution of scattering clouds would be
less effective at attenuating the radiation because the threshold would be
harder
to satisfy), then we can roughly describe the scattered UV and X-ray radiation
by
an isotropic radiation field. Equation (18) can be approximated by

$$\tau_{\gamma-\gamma}(\epsilon_1) \cong \int_{z_i}^z dz'\int_{{2\over
\epsilon_1}}^\infty d\epsilon\;\sigma_{\gamma-\gamma}(\epsilon_1,\epsilon)
n_{ph}(\epsilon;z')\; . \eqno(19)$$

\noindent We use the convenient approximation

$$\sigma_{\gamma-\gamma}(\epsilon_1,\epsilon) \cong {1\over
3}\;\sigma_T\epsilon\delta(\epsilon-{2\over\epsilon_1}) \eqno(20)$$

\noindent for the pair-attenuation cross section (Zdziarski \&
Lightman 1985), after correcting the misleading notation.

Assume that the central source produces total radiant luminosity $L =
10^{46}L_{46}$ ergs s$^{-1}$ and spectral luminosity $L(\epsilon) =
L_o\epsilon^{1-\alpha}$ in the 1-10 keV band. This energy band is most
effective
at attenuating $\approx$100 MeV - 1 GeV gamma rays. If $\alpha = 1.7$, it is
easy
to show that

$$n_{ph}(\epsilon;z') \approx {L(\epsilon)\tau_{sc}(z')\over 4\pi z'^2c\epsilon
m_ec^2} \cong 6.5\times 10^{38}\tau_{-2}(z')\;{L_{46}\epsilon^{-1.7}\over
z'^2}\eqno(21)$$

\noindent (compare eq.[10]). Defining $E_{GeV} \cong\epsilon_1/2000$ and
substituting equations (20) and (21) into equation (19), we obtain

$$\tau_{\gamma-\gamma}(E_{GeV};z)\cong  1.8\times
10^{18}E_{GeV}^{0.7}L_{46}\int_{z_i}^z dz'\;{\tau(z')\over z'^2}.\eqno(22)$$

\noindent The central source emission is scattered by electrons with scattering
optical depth $\tau(z')\ll 1$ according to the relation

$$\tau(z) = \sigma_T\int_0^z dz'\;n_e(z')\;,\eqno(23)$$

\noindent where $n_e(z')$ denotes the radial density
distribution of electrons. If  we assume that $n_e(z') = n_e^o$, a constant,
and
let $n_e^o\sigma_T z = 0.01\tau_{-2}$ at z = 0.1 pc, equation (22) implies

$$\tau_{\gamma-\gamma}(E_{GeV};z)\cong 0.06 E_{GeV}^{0.7}
L_{46}\tau_{-2}\ln({z\over z_i})\;.\eqno(24)$$

\noindent For these parameters, pair attenuation is not significant for GeV
photons.

If we make the unphysical assumption that $\tau(z) = 0.01\tau_{-2}$,
independent
of z, then we find that $\tau_{\gamma-\gamma} = 1$ at

$$z_i (\rm{pc})\lte 0.006E_{GeV}^{0.7}L_{46}\;\tau_{-2}\;.\eqno(25)$$

\noindent This still shows that pair attenuation is considerably less
important than estimated by Blandford.

The difference between our result and Blandford's estimate for 3C 279 stems
partly
from his approximation for $\tau(z)$ and primarily from his assumption
for the soft photon luminosity. He assumes a
1-10 keV luminosity $\approx 2\times 10^{47}$ ergs s$^{-1}$, which
approximately
matches the Ginga observations in the high state (Makino et al. 1989). The
rapid
variability of the X-rays and the failure to detect a UV bump imply, however,
that
the X-rays are probably beamed emission from the jet, so that the level of
isotropic emission must be considerably smaller. Contemporaneous Ginga
observations (Makino et al. 1993) reportedly show that the 2-10 keV X-ray
energy
flux measured early in the gamma-ray flare (June 17-18, 1991) is $1.68\times
10^{-11}$ ergs cm$^{-2}$ s$^{-1}$, implying a 2-10 keV luminosity $\sim
10^{46}$
ergs s$^{-1}$ (H$_o$ = 75 km s$^{-1}$ Mpc$^{-1}$). Combined with the evidence
for
the lack of significant X-ray absorption in the line-of-sight, we see that the
location of the gamma-ray emission site during this time interval is apparently
limited primarily by pair attenuation with accretion-disk photons rather than
scattered photons. A stronger conclusion regarding the location of the
gamma-ray
emission sites will require additional contemporaneous X-ray and gamma-ray
observations.

\section{DISCUSSION}

There is no compelling evidence that the 100 MeV - 1 GeV gamma rays from
blazars
are made farther than $\approx 10^2-10^3 R_g$ from the central engine. This
considerably reduces demands on models of particle acceleration in gamma-ray
blazars; for example, the gamma-ray flaring of 3C 279 by a factor of 400\% in a
two-day period (Kniffen et al. 1993) would seem to be difficult to explain
(even
with beaming) if the emission site were located $\approx$ one-light month from
the
black hole.

The Compton-scattering energy loss rate of relativistic electrons in outflowing
plasma blobs is dominated by accretion-disk photons rather than scattered
photons
at distances $\lte 0.01-0.1$ pc from the central source, and even farther if
the
source lacks scattering clouds. This eliminates any need to invoke separate
source
models for quasars and BL Lac objects on the basis of the standard AGN scenario
which views BL Lac objects as weak quasars which lack emission line clouds; of
course, the true situation is undoubtedly more complicated, with a gradation of
scattering-cloud column densities.

Our estimate (24) for the pair attenuation depth implies that TeV photons will
nonetheless be strongly attenuated ($\tau_{\gamma-\gamma}\approx 10$) by
scattered
radiation for the stated parameters. Indeed, no quasars have been detected with
the Whipple observatory (Fennell et al. 1992), but a BL Lac object, Mrk 421,
has
been detected (Punch et al. 1992). This is in accord with the AGN unification
scenario if accretion-disk photons instead of scattered photons are the primary
source of radiation which is Compton-scattered by the jet.  An important
implication follows: in order to avoid attenuation by intergalactic radiation
fields, the most likely sources of extragalactic TeV radiation are nearby (z
\lte
0.1-0.2) sources (Stecker, de Jager, \& Salamon, \& 1992). But they are less
likely to be detected from nearby quasars (such as 3C 273) or FR II radio
galaxies
because of the presence of emission line clouds. Rather, TeV gamma rays are
more
likely to be seen from nearby lineless BL Lac objects.

 We thank F. Stecker and the referee for useful comments.  R. S. acknowledges
partial support by the DARA (50 OR 93011 of his {\it GRO} guest investigator
program GRO-90-44.
\pagebreak[4]

\end{document}